# Resistance noise scaling in a two-dimensional hole system in GaAs

R. Leturcq[a], G. Deville[a], D. L'Hôte*[a], R. Tourbot[a], C.J. Mellor[b], M. Henini[b]

[a]Service de Physique de l'Etat Condensé, DSM, CEA- Saclay, F-91191 Gif-sur-Yvette Cedex, France ;
[b]School of Physics and Astronomy, University of Nottingham, University Park, Nottingham NG7 2RD, United Kingdom

## ABSTRACT

The $1/f$ resistance noise of a two-dimensional (2D) hole system in a high mobility GaAs quantum well has been measured on both sides of the 2D metal-insulator transition (MIT) at zero magnetic field ($B = 0$), and deep in the insulating regime. The two measurement methods used are described: $I$ or $V$ fixed, and measurement of resp. $V$ or $I$ fluctuations. The normalized noise magnitude $S_R/R^2$ increases strongly when the hole density is decreased, and its temperature ($T$) dependence goes from a slight increase with $T$ at the largest densities, to a strong decrease at low density. We find that the noise magnitude scales with the resistance, $S_R/R^2 \sim R^{2.4}$. Such a scaling is expected for a second order phase transition or a percolation transition. The possible presence of such a transition is investigated by studying the dependence of the conductivity as a function of the density. This dependence is consistent with a critical behavior close to a critical density $p^*$ lower than the usual MIT critical density $p_c$.

Keywords: Two-dimensional systems, 1/f noise, phase transitions, percolation

## 1. INTRODUCTION

Two-dimensional (2D) electronic systems at low temperature offer the unique possibility of studying the interplay between disorder and interactions which underlies the physical properties of electron gases in condensed matter. At low density, the standard description of non-interacting electrons is dominated by the crossover from weak localization (WL) to strong localization (SL). In this case, the disorder is the major ingredient: its magnitude determines the density at which the transport of independent particles will go from diffusive (in the WL regime) to hopping (in the SL regime). However, if the disorder is very low, the strength of the Coulomb interactions can become so large when the density is decreased that the SL should not be reached because the particles should no longer be considered as independent. In this case, a new physics related to the strong interactions between the electrons is expected. It should appear when the correlations between the carriers overcome the random motion of the electrons due to their confinement. The relative magnitude of these two effects is given by the ratio $r_s = E_{ee}/E_F$ between the Coulomb interaction ($E_{ee}$) and the Fermi ($E_F$) energies. $r_s$ is proportional to $m^*/p_s^{1/2}$, $m^*$ being the effective mass of the carriers, and $p_s$ their areal density. For $r_s \gg 1$, one expects to observe a Wigner crystal[1], thus raising the question of the nature of the transition to this state by varying the density[2,3].

The observation of a metallic behavior at intermediate $r_s$ values, $4 < r_s < 36$, in 2D electron or hole systems (2DES or 2DHS) in high mobility silicon metal-oxide-semiconductor field effect transistors (Si-MOSFETs) and in certain heterostructures has raised the possibility of a new metallic phase due to the interactions[4-7], in contradiction with the scaling theory of localization for independent particles[8]. The metallic behavior is defined by a decrease of the resistivity $\rho$ for decreasing temperature $T$, for $p_s > p_c$ where $p_c$ is a critical density. When $p_s < p_c$, an insulating behavior occurs ($d\rho/dT < 0$). This possible MIT has been intensively studied, and strongly debated[3-7,9-54]. At the present time, explanations of the metallic behavior by corrections to the standard independent particle picture, such as temperature-dependent interaction and screening effects[26-38] have been put forward. Meanwhile, the question of the physical nature of a correlated system at large $r_s$ remains open.

Real systems are subject to disorder. Experimentally, the physical observables depend significantly on the disorder[4,13-18]. Weak disorder could reduce the $r_s$ threshold for Wigner crystallisation[55]. For a stronger disorder, the system may also freeze into a glass instead of crystallizing[17,18,51-53,56-59], as recently found in Si-MOSFETs[17-20]. In GaAs 2DHS, local electrostatic studies[22,23] and transport measurements in a parallel magnetic field[25] suggest the coexistence of two phases.

*lhote@drecam.saclay.cea.fr

Such a situation has been predicted by theories in which the interactions and/or the disorder lead to a spatial separation of a low and a high density phase[3,43,44,46-48]. The Wigner crystal (or glass) and the Fermi liquid could be separated by various intermediate phases[3]. In such theories, the transport and other physical properties could be due to the percolation of one of the phases through the other, each of them being conducting or insulating. Another (different) percolation description of the system consists in Fermi liquid puddles connected by quantum point contacts[39-41]. In the present study, we use $1/f$ noise as a tool to investigate the physical properties of a 2DHS in p-GaAs.

## 2. METHODOLOGY

### 2.1 Samples

The two-dimensional hole systems (2DHSs) are created in Si modulation doped (311)A high mobility GaAs quantum wells. The gate used to change the density is a NiCr layer evaporated onto a 1 μm thick pyralin film. This insulating polymide reduces the gate-2DHS current to a negligible level, thus both the transport and noise measurements are unaffected by any leakage from the gate. The large distance between the gate and the 2DHS makes the gate screening of the hole-hole (h-h) interaction negligible, the average distance between neighbouring holes being of the order of 80 nm in the density range we considered. The experiments were carried out on Hall bars 50 μm wide and 300 μm long. The mobility is $5.5\times10^5$ cm$^2$/V.s at a density $p_s = 6\times10^{10}$ cm$^{-2}$ and a temperature $T = 100$ mK. Such a large value guarantees the very low disordered nature of the potential landscape seen by the holes. The "clean" nature of the 2DHS is confirmed by the temperature dependence of the resistivity $\rho$ shown on Fig. 1(a) and the well defined plateaus and oscillations of the Hall and Shubnikov-de-Haas curves at a density so low as $3.8\times10^{10}$ cm$^{-2}$ (see Fig. 1(b)). The $\rho(T)$ increase when $T$ goes from 40 to 600-1000 mK is by a factor almost 2, comparable to the values obtained in other experiments in high mobility p-GaAs[4,10-12].

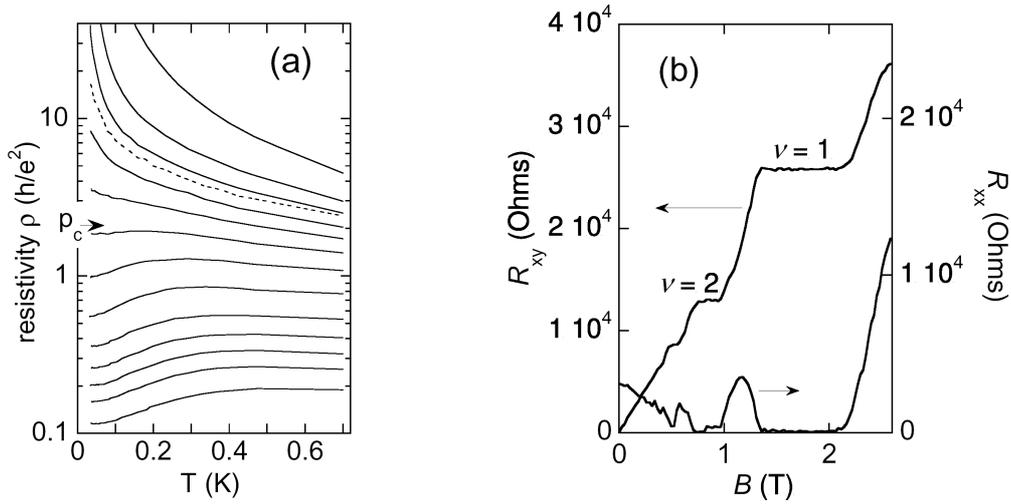

Fig. 1. (a) Longitudinal resistivity $\rho$ as a function of the temperature for densities $p_s = $ 1.30, 1.39, 1.43, 1.44, 1.48, 1.52, 1.57, 1.63, 1.74, 1.86, 1.96, 2.06, 2.16, 2.31 $\times 10^{10}$ cm$^{-2}$ (from top to bottom). The dashed line is the $\rho(T)$ curve corresponding to the $p_s$ limit $p'_c = 1.44 \times 10^{10}$ cm$^{-2}$, under which the activated law analysis is valid. (b) Longitudinal and transverse resistance as a function of the magnetic field at a density $p_s = 3.8 \times 10^{10}$ cm$^{-2}$ and T = 150 mK.

### 2.2 Experimental methods

We performed transport and resistance noise measurements at densities ranging from about $1.4\times10^{10}$ cm$^{-2}$ to $1.8\times10^{10}$ cm$^{-2}$ and temperatures from 35 to 700 mK. The samples are glued on a copper plate thermally connected to the mixing chamber of a dilution refrigerator. The resistor used as a heater, and the RuO$_2$ thermometer are fixed inside the copper plate. The sample and sample holder are placed in a metallic box internally covered with a coating which absorbs IR radiations. The design of the box is such that it avoids electromagnetic radiations emitted by "hot" parts of the cryostat from reaching the sample. Such radiations are known to affect the ionization state of impurities in semiconductor structures placed at very low temperature[60]. The thermal impedance between the sample holder and the mixing chamber is such that sample temperatures between 30 mK and 2 K can be obtained without affecting the dilution process of the refrigerator. A special attention has been put in the protection against parasitic contributions to the noise. The temperature was fixed by injecting a given power in the heating resistor. We verified that this method did not induce measurable temperature fluctuations, contrary to our commercial temperature regulator using the PID method. The RF shielding concerned not only the sample, its wiring, etc, but also the low noise preamplifiers. It was improved by using RF absorbing foam. To minimize microphonic effects, the whole cryostat is placed on laminar flow vibration isolators. The whole experiment is installed on a concrete slab lying on sand in order to avoid the transmission of the low frequency vibrations in the building to the experimental device. To uncouple the cryostat from external vibrations we used absorption and dissipative materials on the cryostat itself and on the pipes connected to the pump and compressor of the dilution circuit.

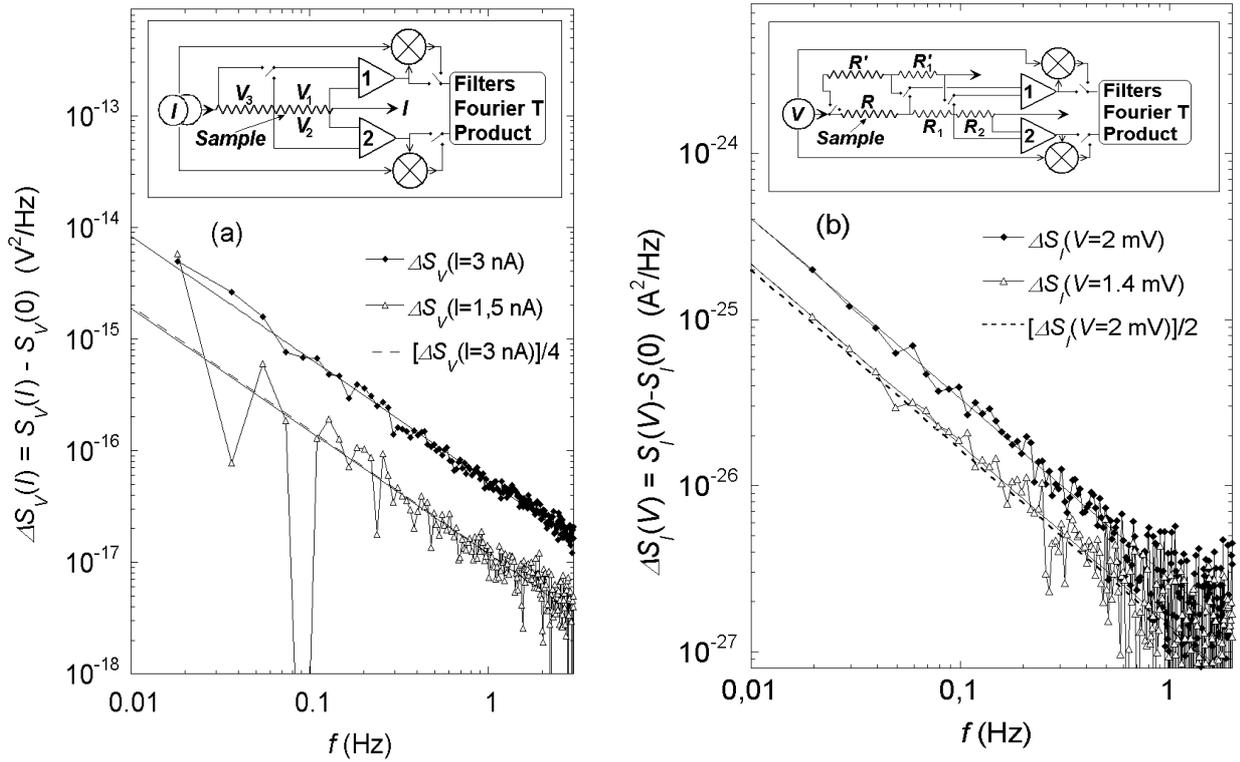

Fig. 2. (a) Comparison of two spectra obtained with the fixed current method, for $I = 1.5$ nA (open triangles) and $I = 3$ nA (closed diamonds). The temperature is 300 mK and the density $1.75\times10^{10}$ cm$^{-2}$. The two $\Delta S_V(I)$ data curves have been fitted with $A/f^\alpha$, $A$ and $\alpha$ being the parameters of the fits (full lines). The dashed line is the $A/f^\alpha$ spectrum for $I = 3$ nA, divided by four. The proportionality of $\Delta S_V(I)$ to the square of the current is verified. Inset of (a): Measurement circuit. The noise power $S_R$ is obtained from the correlation between $V_1(t)$ and $V_2(t)$. The weakness of the correlation between $V_2(t)$ and $V_3(t)$ is a verification that the measured voltage fluctuations are not due to fluctuations of $I$, $T$, $V_G$, etc. The possible use of a modulated excitation $I$ is indicated. (b) Comparison of two spectra obtained with the fixed voltage method, for $V = 1.4$ mV (open triangles) and $V = 2$ mV (closed diamonds). The temperature is 700 mK and the gate voltage $V_G = 2.238$ V. The two $\Delta S_I(V)$ curves have been fitted with $A/f^\alpha$ (full lines), the dashed line is the $A/f^\alpha$ spectrum for $V = 2$ mV, divided by 2.04. Inset of (b): Measurement circuit. The current noise power $S_I$ is obtained from the correlation of the voltages across the two resistors $R_1$ and $R_2$. A possible way to verify that the current fluctuations are not due to fluctuations of $V$, $T$, $V_G$, etc is indicated: it consists in using the correlation between the voltages across $R_2$ and $R'_1$, $R'$ being a second sample with the same gate voltage. The possible use of a modulated excitation $V$ is indicated.

Depending on the sample resistance, two measurement methods were used. For resistances below $10^6$-$10^7$ Ohms, the voltage and its time fluctuations were measured, for a fixed injected current $I$ (see inset of Fig. 2(a)); while for resistances above $10^6$-$10^7$ Ohms, the current and its time fluctuations were measured, for a fixed applied voltage $V$ (see inset of Fig. 2(b)). $I$ or $V$ were chosen low enough to ensure Ohmic conduction. Both DC and AC techniques were used. The typical frequency interval in which the noise power spectrum was recorded is 0.01 Hz $< f <$ 3 Hz. When the AC technique was used, only the low frequency part of the spectra ($f <$ 1 Hz) was recorded as the high frequency part was affected by the 2 to 10 Hz modulation signal. In what follows, we present the normalized resistance noise power $S_R/R^2$, $R$ being the average value of the 2DHS in the Ohmic region.

In the fixed current method, the voltage noise power spectrum $S_V$ was obtained by using the cross-correlation technique[61] for the the two voltage noise signals measured on opposite sides of a Hall bar ($V_1$ and $V_2$ on inset of Fig. 2(a)). As a result, the contribution of the noise due to the contacts, leads and preamplifiers is suppressed. The resistance noise is obtained from the equation $S_R = [S_V(I) - S_V(0)]/I^2$. In the current interval where the conduction is Ohmic, we verified that $[S_V(I) - S_V(0)]/I^2$ did not depend on $I$ (see Fig. 2(a)), as expected if $S_V(I) - S_V(0)$ comes only from resistance noise. To verify that possible spurious noise sources (e.g. fluctuations of $I$, $T$, gate voltage $V_G$, etc.) did not contribute to $S_V(I) - S_V(0)$, we measured the power spectra $S'_V$ corresponding to the correlation between the voltage noise along two contiguous sections of the Hall bar ($V_2$ and $V_3$ on inset of Fig. 2(a)). $S'_V(I)$ and $S'_V(0)$ had a flat frequency dependence and their magnitudes were close to $S_V(0)$ and well below $S_V(I)$, thus confirming that only the sample contributes to $S_V(I) - S_V(0)$.

In the fixed voltage method, the current noise power spectrum $S_I$ was obtained by using the cross-correlation technique for the two voltage noise signals measured on two resistors in series with the sample ($R_1$ and $R_2$ on inset of Fig. 2(b)). The resistance noise is obtained from the equation $S_R = (R+R_1+R_2)^4 \cdot [S_I(V) - S_I(0)]/V^2$, $S_I$ being obtained as the correlated part of the two measured voltages, divided by $R_1 R_2$. The two resistors $R_1$ and $R_2$ were low noise devices chosen in order that their 1/$f$ noise gives a negligible contribution to $S_I$. We verified that $[S_I(V) - S_I(0)]/V^2$ was independent of $V$ in the Ohmic conduction interval. The verification that possible contributions of $V$, $T$, gate voltage fluctuations do not contribute to $S_I(V) - S_I(0)$ can be made by looking at the correlation between the voltage fluctuations across $R_2$ and $R'_1$ in the inset of Fig. 2(b); $R'_1$ being in series with a second sample with the same gate voltage. We verified that the four different methods (fixed $I$ or $V$, AC or DC) gave the same results.

## 3. EXPERIMENTAL RESULTS

### 3.1 DC transport measurements

The $T$-dependences of the 2DHS resistivity $\rho$ at various densities on Fig. 1(a) show a metallic behavior at the largest densities. It consists in an increase of $\rho$ with the temperature. On the contrary, at the lowest densities, $\rho$ seems to diverge when $T \rightarrow 0$. The change of slope from d$\rho$/d$T > 0$ to d$\rho$/d$T < 0$ attributed to the 2D MIT[4] occurs at $p_c = (1.57 \pm 0.02) \times 10^{10}$ cm$^{-2}$. Assuming an effective mass $m^* = 0.37\, m_e$[62], the value of $r_s$ corresponding to $p_c$ is $r_{sc} \approx 24$. Because of this high value, strong interaction effects can be expected. At the lowest densities and temperatures, i.e. for $T$ below a limit $T_l(p_s)$, we find $\rho(T) \propto \exp(T_0/T)$, as found in Si-MOSFETs[17,18,21,63]. $T_0$ depends linearly on $p_s$ and vanishes at $p_c' = (1.44 \pm 0.05) \times 10^{10}$ cm$^{-2}$. Clearly, this temperature dependence contrasts with the variable range hopping (VRH) transport laws characterizing the SL regime[64].

### 3.2 1/$f$ noise measurements

The density interval in which the resistance noise spectra have been obtained goes from about 1.40 to $1.78 \times 10^{10}$ cm$^{-2}$, thus on both sides of the MIT ($p_c \approx 1.57 \times 10^{10}$ cm$^{-2}$). The data at the largest densities[49-50] have been completed by measurements at $p_s \leq 1.50 \times 10^{10}$ cm$^{-2}$. Fig. 3 gives an example of the resistance fluctuations at low density. The spectra could be fitted with a law $S_R(f) = A/f^{\alpha}$, $A$ and $\alpha$ being the fit parameters. $\alpha$ does not show any strong variation as a function of $p_s$ or $T$, and remains in the interval 0.9-1.4. We do not find a saturation of the noise at low frequency. The normalized noise magnitude at 1 Hz, $S_R/R^2$ (1 Hz) $= A/R^2$ is shown on Fig. 4 for various densities ranging from 1.5 to $1.78 \times 10^{10}$ cm$^{-2}$. It increases strongly when $p_s$ decreases, while the temperature dependence goes from a weak increase with $T$ at the largest densities to a decrease that becomes steeper and steeper as $p_s$ decreases (see Fig. 4(b)).

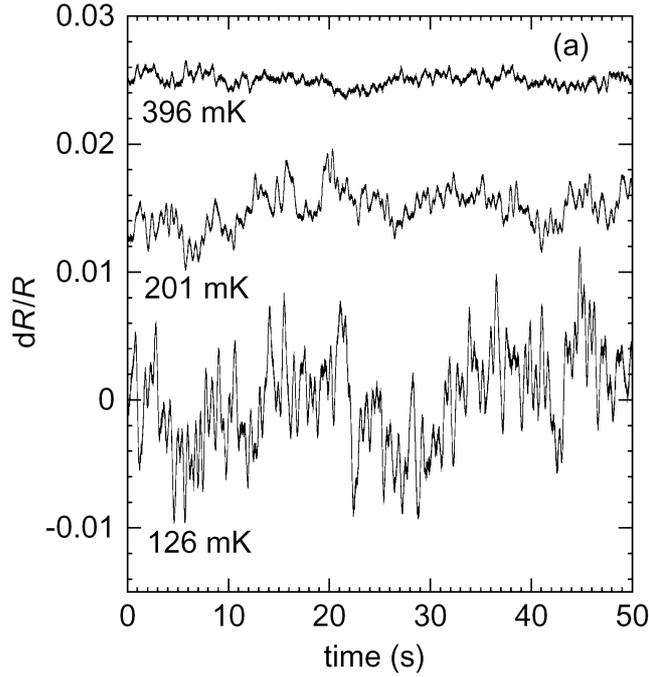

Fig. 3. Typical relative fluctuations of $R$, $dR/R = (R(t)-\langle R\rangle)/\langle R\rangle$, as a function of time for three different temperatures, at a gate voltage $V_G = 2.23$ V. To keep only the fluctuations of interest, a low-pass filter has been used, with a corner frequency $f_c = 2$ Hz. The density is so low ($p_s < 1.45\times10^{10}$ cm$^{-2}$), that the cross-correlation technique is not needed in order to get rid of unphysical contributions to the noise. The traces have been shifted for clarity.

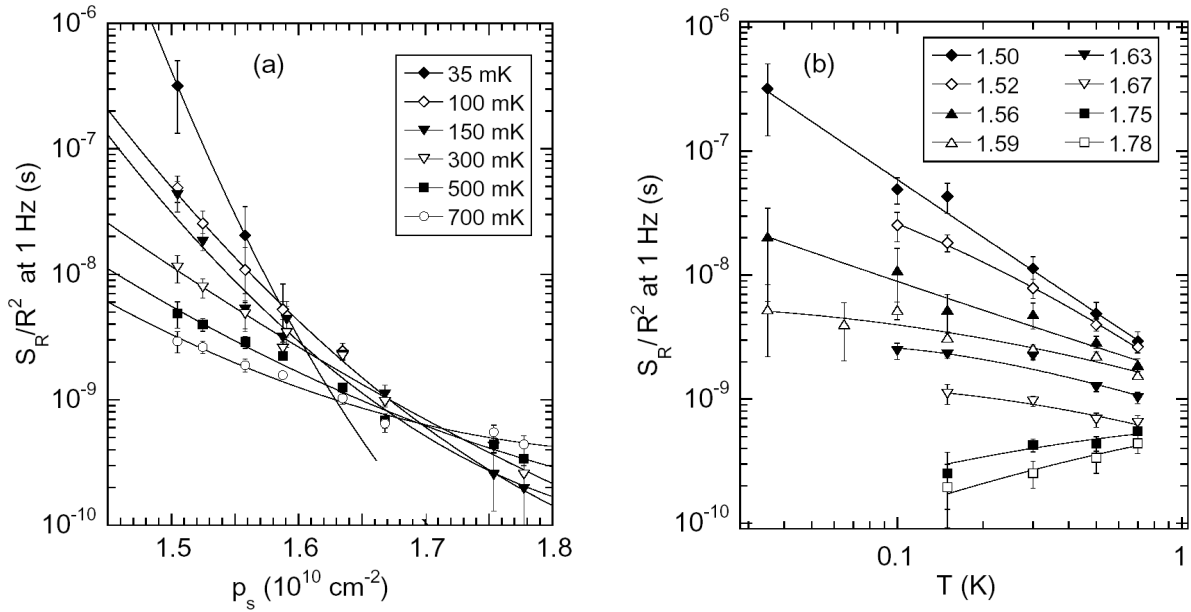

Fig. 4. (a) Normalized resistance noise power $S_R / R^2$ at 1 Hz, as a function of the density $p_s$ for seven temperatures. (b) The same results plotted as a function of the temperature for eight densities indicated in units of $10^{10}$ cm$^{-2}$. In (a) and (b), the lines are guides to the eye.

The temperature dependences measured at densities lower than $1.5 \times 10^{10}$ cm$^{-2}$ have been obtained by using the fixed voltage method mentioned above. Figure 5 gives preliminary results indicating that the temperature dependence continues to become steeper as the density decreases. The average slope of the $\ln(S_R/R^2)$ vs. $\ln(T)$ dependence in the $T =$ 100-500 mK interval goes from ~ -0.5 to ~ -1.5 when the density decreases from 1.6 to $1.5 \times 10^{10}$ cm$^{-2}$, (see Fig. 4(b)). The same average logarithmic slope goes to values lower than -3 for the data displayed on Fig. 5.

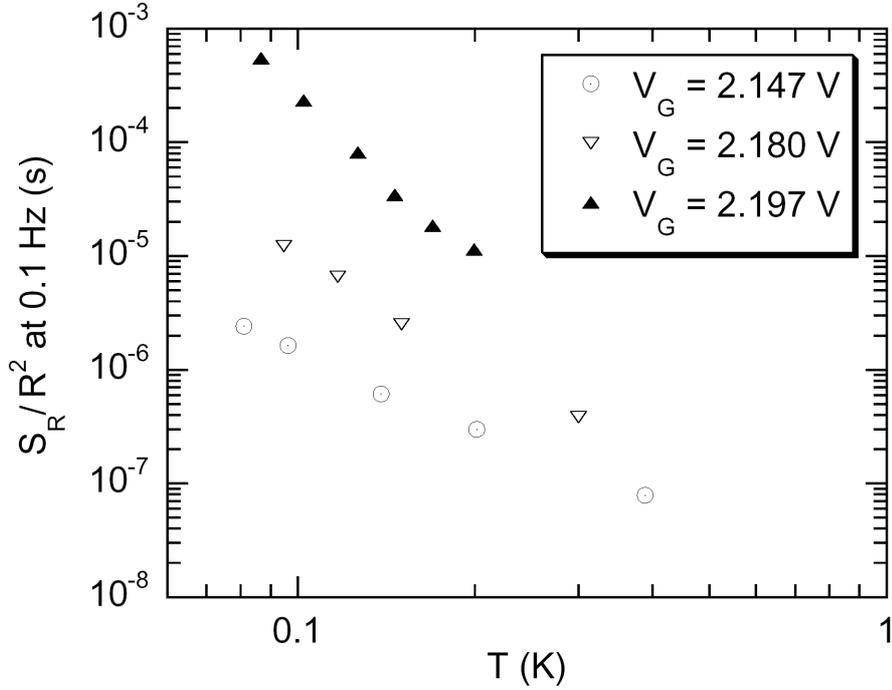

Fig. 5. Normalized resistance noise power $S_R/R^2$ at 0.1 Hz, as a function of the temperature for three values of the gate voltage $V_G$ used to control the density (preliminary results). The corresponding estimated densities lie in the interval 1.4 - $1.5 \times 10^{10}$ cm$^{-2}$ (the density is a decreasing function of $V_G$).

The strong increase of $S_R/R^2$ when $p_s$ decreases (see Fig. 4(a)), and the evolution of the temperature dependence suggests that a new regime occurs at low density. The observed temperature dependences of the noise in the insulating regime ($p_s < 1.57 \times 10^{10}$ cm$^{-2}$), and the absence of saturation at low frequency can hardly be explained by existing noise models based on the hopping of electrons[65-73]. At our largest densities, the relevance of noise models which assume independent particles in the diffusive regime must be addressed. A decrease of the noise magnitude when the temperature increases is expected for a degenerate system at low temperature because of quantum interference effects[74-75]. However, for simple diffusion in the non-degenerate case, the thermal fluctuations of scattering centers[76] leads to an increase of $S_R/R^2$ with the temperature. In the Dutta-Horn model[77], the temperature dependence is $S_R/R^2 \propto T$ for $1/f$ spectra. The slopes we obtain at $p_s = 1.75\text{-}1.78 \times 10^{10}$ cm$^{-2}$ are $\partial \ln(S_R/R^2)/\partial \ln(T) \approx 0.5$. However, the distribution of energy levels in our samples could be outside the model assumptions. The increase we observe at high density is of particular interest because it could be related to the corrections to Drude transport such as interaction effects and temperature dependent screening[26-38]. Such effects have been put forward as the explanation of the positive d$\rho$/d$T$.

The strong increase of the noise when $T$ decreases has been observed in Si-MOSFETs[17-20], and analyzed as due to the freezing of a glass at low density. However, we do not observe an increase of the parameter $\alpha$ ( of the $S_R(f) \sim 1/f^{\alpha}$ dependence) from 1 to 1.8 when $p_s$ decreases, as in Si-MOSFETs. This difference could be due to the low level and the different nature of the disorder in $p$-GaAs in comparison to Si-MOSFETs. Many studies have probed the physical importance of the disorder magnitude[4,13-18], but its nature could play a role too[30,31,38,43,44,48]. For instance, its spatial range could govern the phase separation[43,44]. We note that specific features of p-GaAs in comparison with Si-MOSFETs, such as the magnitude and nature of the disorder must be considered to explain the notable differences in the $\rho(T, p_s)$ dependences. The order of magnitude separating the critical densities $p_c$ is one of them. It should be also noted that the frequencies at which our spectra were recorded are larger than those of Refs. 17-20.

A striking feature of our data is that when the points shown on Fig. 4 ($1.50 \times 10^{10}$ cm$^{-2}$ ≤ $p_s$ ≤ $1.78 \times 10^{10}$ cm$^{-2}$) are plotted as a function of the resistance $R$, they collapse on a single curve which appears to be compatible with a power law (see Fig. 6(a)). A fit of all the points gives $S_R/R^2 \propto R^{2.40 \pm 0.06}$. Fig. 6(b) shows that the power law dependence remains true at a given temperature. This scaling suggests the existence of a second order phase transition which would occur at a critical density $p^*$ lower than $1.50 \times 10^{10}$ cm$^{-2}$. It could be a transition separating two phases of the system[2-4], or a percolation transition as suggested by various models evoked above[3,4,39-41,43-44,45-48].

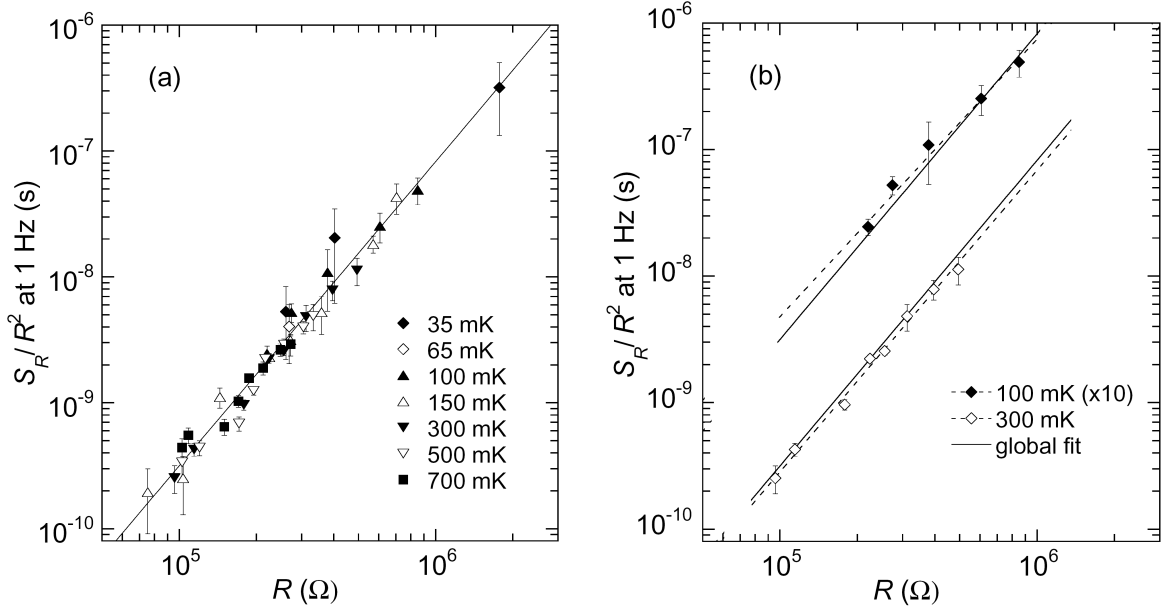

Fig. 6. (a) Normalized resistance noise power $S_R/R^2$ at 1 Hz as a function of the resistance, for the eight densities of Fig. 4, and for seven temperatures. The line is a fit with a power law $S_R/R^2 \propto R^{2.4}$. (b) Same plot as (a), except that only two temperatures have been selected: 100 and 300 mK. For clarity, the data and fits at 100 mK have been shifted by a factor 10. The two dashed lines are two separate fits of a power law on the data at each temperature. The full lines correspond to the global fit presented in (a).

As a matter of fact, the origin of the observed scaling could be that both the conductance $\sigma = 1/R$ and the noise follow scaling laws $\sigma \propto (x-x^*)^t$ and $S_R/R^2 \propto (x-x^*)^{-\kappa}$ where $x$ would be a physical quantity governing the proximity to a critical point at $x = x^*$, $t$ and $\kappa$ being critical exponents. As a result, $S_R/R^2 \propto R^w$, with $w = \kappa/t$. In the case of a percolation transition, $x$ is the fraction of connected bonds between the sites.

Many experimental and theoretical works have been devoted to the study of the critical exponents for the percolation of a conducting network in an insulating matrix[78-84]. While $t = 1.3$ is expected in two dimensions in most cases, the value of $w$ depends on the nature of the network. This fact stresses the interest of noise studies for probing the physics of a system. 2D simulations give $w = 1$ for a square lattice[79], and $w = 3.2$ for a random-void network[80]. It is reasonable to assume that in our case, a conducting phase percolating in an insulating one (or a network of Fermi liquid puddles connected by quantum point contacts) is in a sense intermediate between those two extreme limits. Other values found experimentally or theoretically go from $w = 2$ to $w = 4.2$[81-83].

## 3.3 Discussion

The relationship between 1/$f$ noise and the resistance has been already studied in three-dimensional systems such as indium oxide films and ZnO accumulation layers close to the Anderson transition[85,86]. The noise was attributed to fluctuations of the system between the metal and the insulator due to the fluctuations of the disorder potential. The authors showed experimentally and explained theoretically that the relative noise magnitude presents an exponential dependence as a function of the resistance. This dependence was related to the exponential dependence of the resistance on the insulating side as a function of the localization length and to the critical divergence of the latter at the transition. Such a dependence is incompatible with our data. We can thus draw the important conclusion that a quantum phase transition similar to the Anderson transition studied in Refs. 85,86 is excluded in our 2DHS.

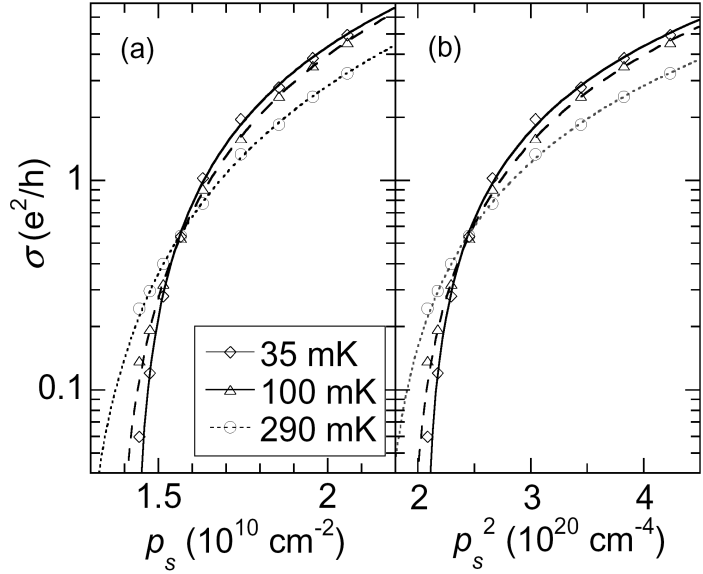

Fig. 7. (a) Conductivity $\sigma$ as a function of the density $p_s$ for three temperatures. The lines are fits with $\sigma = \lambda [p_s - p^*(T)]^t$, $\lambda$, $p^*$ and $t$ being parameters of the fits. (b) $\sigma$ vs. $p_s^2$, and the lines are fits with $\sigma = \lambda [p_s^2 - p^{*2}(T)]^t$.

The $S_R/R^2$ vs $R$ dependence shown on Fig. 6 can be explained by a critical behavior close to a transition leading to the scaling of those two observables as a function of $x - x^*$ as considered above. However as another mechanism could lead to the same behavior, it is interesting to investigate the possible scaling of the conductivity, i.e. $\sigma \propto (x - x^*)^t$. The fact that our experimental points in Fig. 6 scale together whatever the temperature suggests that $x$ is a function of both $p_s$ and $T$. The simplest relationship $x \propto p_s$ is obtained in a percolation model where the disordered potential landscape is progressively filled by the Fermi liquid when the density increases[39-41]. In such a description, the transport occurs by percolation through a network of Fermi liquid puddles connected by quantum point contacts. We may, after having considered the $x \propto p_s$ case, study a more complex dependence $x \propto p_s^\beta$. Such a law with $\beta > 1$ is suggested in a case where the system is made of a conducting phase percolating in an insulating one. Various models, in which the interactions play the major role, predict such a separation in two phases[3,4,39-41,43-44,45-48].

To investigate the $x \propto p_s$ case, we fitted the $\sigma$ vs. $p_s$ dependences for $p_s > 1.50 \times 10^{10}$ cm$^{-2}$ with the law $\sigma = \lambda [p_s - p^*(T)]^t$, $\lambda$, $p^*$ and $t$ being parameters of the fits (see Fig. 7(a)). The fits are good, but the exponent $t$ varies between 1.35 $\pm$ 0.1 (at 35 mK) and 1.9 $\pm$ 0.2 when the temperature is varied, while a value close to 1.3 is expected in 2D systems[39,40]. Note that a $\sigma \propto (p_s - p_0)^{1.3}$ dependence had been found in 2DHS's similar to ours[39,40]. Fig. 7(b) shows the case $x \propto p_s^2$. Again the fits are good but the $t$ values are closer to each other and to 1.3 when $T$ varies: they range from 1.2 $\pm$ 0.1 to 1.45 $\pm$ 0.15. This kind of study can be performed for any value of $\beta$. Figure 8 (left) shows the values of $t$ extracted from fits with $\sigma = \lambda [p_s^\beta - p^{*\beta}(T)]^t$ where $\beta$ is fixed. The main conclusion of these $\sigma$ vs. $p_s$ studies is that they are compatible with the possibility that the noise vs. resistance law results from a critical behavior of both $S_R/R^2$ and $\sigma$ close to a critical density $p^*$.

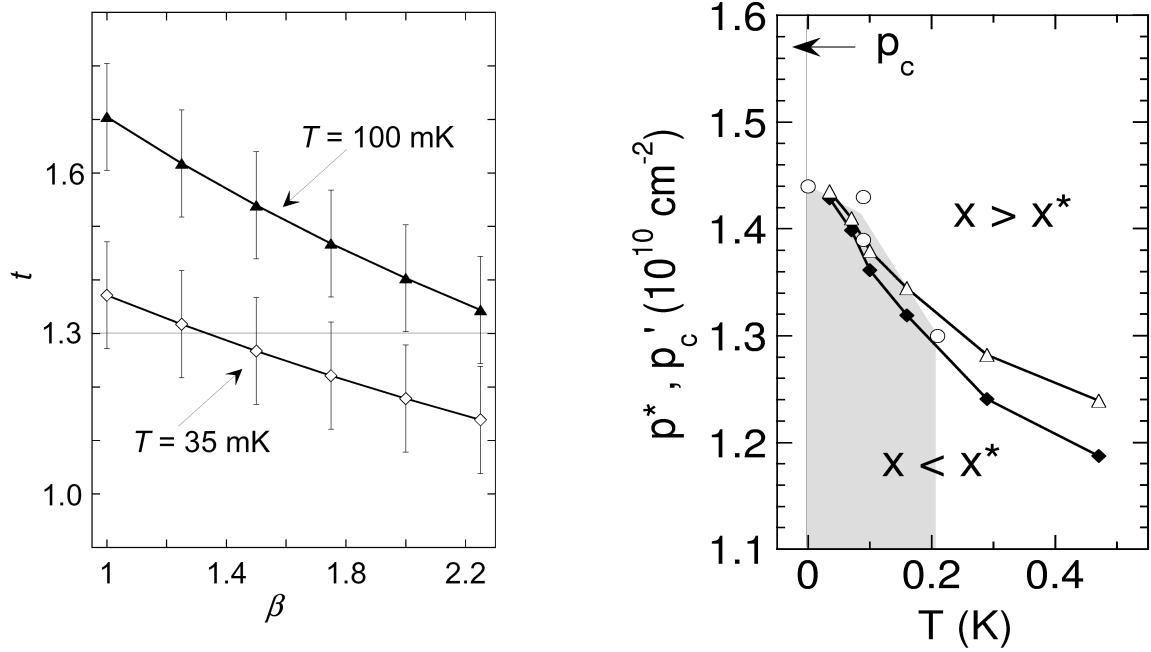

Fig. 8. Left: The exponent $t$ extracted from the fit of the data $\sigma(p_s)$ (obtained at two temperatures 35 and 100 mK) with the law $\sigma = \lambda [p_s^\beta - p^{*\beta}(T)]^t$, as a function of $\beta$ which is a fixed parameter of the fit. The error bars indicate estimated uncertainties resulting from the uncertainties on $\sigma$ and $p_s$. Right: Temperature dependence of the density $p^*$ extracted from the fits shown on Fig. 7. The closed diamonds (open triangles) correspond to $\sigma \propto (p_s - p^*)$ (resp. $\sigma \propto (p_s^2 - p^{*2})$). Open circles: $p_s$ limit below which $R(T)$ can be fitted with an activation law $\propto \exp(T_0/T)$. At $T = 0$, this corresponds to $p_s = p'_c$ (see section 3.1). The shaded area is the domain where the activation law is valid.

Fig. 8 (right) gives the temperature dependence of the density $p^*(T)$ extracted from the fits of $\sigma(p_s)$ for $\beta = 1$ (closed diamonds) and $\beta = 2$ (open triangles). They are close to each other and to the boundary (open circles) which separates the two domains where the activation law analysis $R \propto \exp(T_0/T)$ is valid or not, i.e. the $T_l(p_s)$ line mentioned above. In a percolation description, this result could be explained by the fact that below the percolation threshold, the system consists of isolated conducting regions between which the conduction electrons jump due to thermal activation. However, we note that activated transport laws have been also proposed for the transport in the case of a pinned Wigner crystal[63,87]. As already stressed in Refs. 39-41, the percolation scenario has the advantage of allowing an explanation of the continuous connexion of the $B = 0$ MIT with the quantum Hall-insulator transition[54]. The percolating nature of the latter has been proposed by several authors[88,89].

## 4. CONCLUSION

By studying resistance noise measurements on a high mobility, gated, $p$-GaAs quantum well at low density, we find a huge increase of the noise power $S_R/R^2$ when the density decreases, which is accompanied by a change of the slope of its temperature dependence. The measurements performed by using the fixed voltage method (measurement of $S_I$) show that this slope becomes steeper at very low density. The noise vs. resistance dependence exhibits a scaling behaviour, $S_R/R^2 \propto R^{2.4}$ for densities above $0.95 \times p_c$ and for all temperatures studied. Such a result suggests the existence of a second order phase transition, or a percolation transition. The conductivity vs. density analyses are compatible with this scenario. They allow the extraction of a critical density $p^*(T)$ which is lower than $p_c$, the usual MIT critical density.

Concerning a percolation transition, two kinds of models must be considered. The first one is that of Fermi liquid puddles connected by quantum point contacts[39-41]. In the second one, the system is a mixture of two phases (conducting and insulating) as suggested by theories of interacting electrons[3,43-48].

## ACKNOWLEDGEMENTS

We are grateful to the Cryogenics Laboratory at SPEC, to P. Pari and P. Forget for their valuable help in building, and improving the dilution refrigerator. We thank J. P. Bouchaud, D. C. Glattli, C. Kikuchi, D. Popović, X. Waintal, F.I.B. Williams and X. C. Xie for useful discussions.